\documentclass{article}

\usepackage[utf8]{inputenc}
\usepackage{graphicx}
\usepackage{a4wide}             
\usepackage{color}              
\usepackage{amsmath}            
\usepackage{amssymb}            
\usepackage{commath}            
\usepackage[english]{babel}
\usepackage{eurosym}
\usepackage{subcaption}         
\usepackage[hyphens]{url}
\usepackage[colorlinks=true,linkcolor=verdoso,citecolor=magenta,urlcolor=azulondos]{hyperref}

\definecolor{azulondos}{rgb}{0,.33,.61}
\definecolor{verdoso}{rgb}{0.1,0.6,0.1}

\title{Self-consumption for energy communities in Spain: a regional analysis under the new legal framework}

\author{Cristobal Gallego-Castillo$^{1,*}$, Miguel Heleno$^2$, Marta Victoria$^{3,4}$ \\
\footnotesize{$^1$DAVE(ETSIAE), Universidad Polit\'ecnica de Madrid, Plaza Cardenal Cisneros 3, 28040 Madrid, Spain.}\\
\footnotesize{$^2$Grid Integration Group, Lawrence Berkeley National Laboratory, One Cyclotron Road, Berkeley, CA 94720, US.} \\
\footnotesize{$^3$Department of Engineering, Aarhus University, Inge Lehmanns Gade 10, 8000 Aarhus C, Denmark.} \\
\footnotesize{$^4$iCLIMATE Interdisciplinary Centre for Climate Change, Aarhus University, Denmark.} \\
\footnotesize{$^*$Corresponding author. Email: \href{mailto:cristobaljose.gallego@upm.es}{cristobaljose.gallego@upm.es} }
}

\begin{document}

\maketitle


\

\section*{Abstract}

European climate polices acknowledge the role that energy communities can play in the energy transition. Self-consumption installations shared among those living in the same building are a good example of such energy communities. In this work, a regional analysis of optimal self-consumption installations under the new legal framework recently passed in Spain is performed.  
Results show that the optimal sizing of the installation leads to economic savings for self-consumers in all the territory, for both options with and without remuneration for energy surplus.
A sensitivity analysis on technology costs revealed that batteries still require noticeably cost reductions to be cost-effective in a behind the meter self-consumption environment. In addition, solar compensation mechanisms make batteries less attractive in a scenario of low PV costs, since feeding PV surplus into the grid, yet less efficient, becomes more cost-effective.
An improvement for the energy surplus remuneration policy in the context of the current legislation was proposed and analysed. 
It consists in the inclusion of the economic value of the avoided power losses in the remuneration.

\

\textit{Keywords}: self-consumption; energy community; residential
buildings; distributed generation; energy transition; Spanish legal framework.

\section*{Nomenclature}

\begin{tabular}{ll}
\textit{AEC}    &   Annualised energy cost \\
\textit{ASR}    &   Annualised savings ratio \\
$C$             &   Average annual electricity consumption per household \\
$C_{\text{\textit{imports}}}$ & Annual cost associated with energy imports from the grid \\
\textit{CC}     &   Capital cost \\
\textit{CP}     &   Energy production cost \\
\textit{EAC}    &   Equivalent annual cost \\
\textit{EIR}    &   Exported-imported ratio \\
$H$             &   Average number of households per building \\
\textit{L}      &   Lifetime \\
\textit{OC}    &   Other costs \\
\textit{OM}     &   Operational and maintenance costs \\
\textit{PERD}   &   Power losses \\
\textit{PMH}    &   Day and intra-day electricity market price \\
\textit{r}      &   Discount rate \\
\textit{SCR}    &   Self-consumption ratio \\
\textit{SSR}    &   Self-sufficiency ratio \\
\textit{TCU}    &   Energy production price \\
\textit{TEU}    &   Energy component of the access tariff \\
$R_{\text{\textit{exports}}}$ & Annual remuneration due to exports to the grid \\
$S$             &   Average building floor area \\ 
\textit{SAH}    &   Adjustment services costs \\
\end{tabular}

\section{Introduction}

Energy communities are key to address the challenge of climate change. The strategy of the European Union (EU) for 2020-2030, defined in the ``Clean energy for all Europeans'' package, acknowledges the need for regulatory frameworks which empowers renewable-based self-consumers (also referred to as \textit{prosumers}) to  generate, consume, store, and sell electricity back to the grid. 
Several advantages can result from the massive deployment of renewable distributed resources (DR) for self-consumption. First, it increases the use of renewable energy sources (RES) in electricity supply, leading to reductions in greenhouse gases emissions. At a technical level, local generation reduces power losses, while deferring future investments in transmission and distribution infrastructure. From the financial point of view, self-consumption deployment entails additional resources stemming mostly from consumers, which helps diversifying centralised energy investments. At the socioeconomic level, distributed generation increases the number of actors that share the benefits associated with the electricity generation activity, historically concentrated in a reduced number of large companies. Additionally, since it is generally accepted that self-consumption will be mostly based on PV systems, it is worth mentioning that distributed PV is associated with higher rates of jobs creation per MW than other energy sources, including large-scale PV \protect{\cite{UKERC2014}}. Together with the aforementioned advantages, photovoltaic self-consumption (PVSC) systems also show some drawbacks. For example, under situations of high PV generation and low electricity demand, a high local concentration of PVSC systems might derive in reverse power flows at the low voltage grid. There is little experience in managing such situations. Also, due to its very nature, installation costs per kW of PVSC systems will always be more expensive than those of large-scale PV installations. This means that, if the socioeconomic impact of having more actors in the sector is disregarded, PVSC is not an optimal solution for PV deployment, as compared with large-scale PV power plants. A comprehensive summary on risks and opportunities of PV self-consupmtion can be found in \cite{LopezProl2020}.

In recent years, the transposition at a national level of European directives related to the Clean energy for all Europeans package translates into new national legal frameworks, including the notion of self-consumption and energy communities \cite{Ines2020}. This implied a radical change in Spain, as previous legislation was a sheer obstacle for self-consumption \cite{LopezProl2017,Mir-Artigues2018}, and poorly rated in many international comparison studies \cite{IEA_review, Gimeno2018}. However, the recent Royal Decree 244/2019 \cite{RD244} ended the taxation of self-consumed energy, introduced remuneration mechanisms for the excess production, and defined the conditions for creating energy communities.

\begin{figure}[ht!]
    \centering
    \includegraphics[width=12cm]{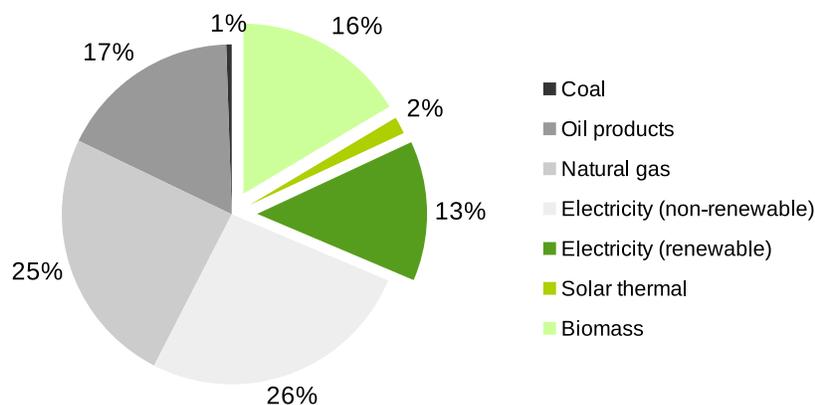}
    \caption{Final energy consumption in the residential sector in Spain (2017), broken down according to energy source. Source: \protect{\cite{IDAE}}.}
    \label{fig_01_energy_consumption} 
\end{figure}

Self-consumption and energy communities allow increasing the renewable penetration in the residential sector. In addition, self-consumption incentives electrification of residential uses, replacing in some cases the local use of natural gas, thus improving  air quality in urban environments. In Spain, the residential sector represented 18\% of the country final energy consumption  in 2017 \cite{IDAE}. Figure \ref{fig_01_energy_consumption} depicts residential final energy consumption according to energy sources, showing that about 69\% originated from non-renewable resources. In addition, only 39\% was covered with electricity. These two facts support the notion that both self-consumption deployment and the electrification of residential uses represent key tools in the design of energy policies.

In this article, PVSC systems in Spain under the new legal framework is analysed, with special focus on energy communities. The aim of this study is to reveal the direct economic impact for self-consumers of the new legislation package. This enables discussing whether support schemes from public institutions are required in order to obtain the aforementioned advantages of self-consumption besides private economic savings. The regional analysis performed in this work provides an answer to this question with higher spatial resolution, which is coherent with the fact that regions in Spain have some degree of autonomy regarding environmental and energy policies, including promotion schemes. Thus, it is important to understand whether or not the different solar radiation and electricity consumption levels among Spanish regions deserve local policies to encompass self-consumption cost-effectiveness at national level. 
Finally, a number of drawbacks of the new legal framework are identified, and specific modifications accompanied by an economic assessment are proposed. 
It is important to stress that the discussion in this paper relates specifically the current self-consumption legislation. Other issues concerning the technical and economical impact of the PV in the system, and cost neutrality of the rate design process need to be addressed in the future.

The rest of this paper is structured as follows. A literature review is performed in Section \ref{sec_review}, revealing some limitations that are addressed in this work. Section \ref{sec_legal_framework} describes the main characteristics of the new legal framework for self-consumption in Spain. This framework is later considered in the simulations and the analysis of this study. Section \ref{sec_data_tool} includes a description of the techno-economical optimisation tool used in this work, together with the different databases considered. In Section \ref{sec_regional_analysis}, results for the optimal self-consumption installation at the regional level are presented, for both cases with and without remuneration for energy surplus. A sensitivity analysis on the relevant economic factors (PV and storage technology costs, discount rate and PV panel lifetime) is conducted in Section \ref{sec_sensitivity}. In Section \ref{sec_remuneration_proposal} the remuneration policy is discussed, and a proposal for improving economic efficiency is analysed. The paper ends with the main conclusions gathered in Section \ref{sec_conclusions}.

\section{Literature review} \label{sec_review}

Self-consumption has been drawing increasingly attention in the scientific community mostly over the last few years. A review on the topic from 2015 concluded that the number of papers was limited at that time, and more comparative studies were needed to assess its potential \cite{Luthander2015}. Since then, many works addressing case studies in different countries and discussing current and potential policies for self-consumption promotion have been published. However, the astonishing reduction in PV and storage costs experienced in the last years, together with the need for increased ambition in energy transition plans for period 2020-2030, call for a continuous update of the research in the field.
In particular, a number of works considering the case study of Spain presented conclusions that are no longer valid, as the legal framework around self-consumption has dramatically changed. 
For example, \cite{Talavera2014} analysed the economic performance of a PV installation at the university of Jaen (southern Spain). Results showed an average Levelised Cost of Energy around 125 \euro/MWh, and a payback time of 17.5 years. 
The analysis performed in \cite{Mir-Artigues2018} considered data from a real case study in Madrid region, consisting in a household PV installation of 2.5 kW, and no storage. The authors concluded that the former regulation was highly unfavourable for the deployment of self-consumption installations, given the small savings obtained.

Three recent papers addressed different aspects of self-consumption in Spain under the new legal framework. 
\cite{Escobar2020} analysed the profitability of self-consumption solar PV systems in Spanish households according to the number of members of the household, from one to four. A comparative with other UE countries (France, Germany, Italy, Great Britain, and Finland) was included, concluding that, even with the improvements of the new legislation, support policies of the other countries were clearly better from an economic perspective. However, this result is partly due to the fact that the research did not consider the possibility of energy excess remuneration in the Spanish case, while current FIT schemes were considered for the other countries. In addition, the study only considers solar radiation data for a single location (Madrid).
\cite{LopezProl2020} assessed the impact of the new PVSC regulation on residential, commercial and industrial prosumers’ profitability, showing that all segments obtain positive profitability in average conditions. In particular, the residential segment had the lowest profitability level.
Finally, \cite{RoldanFernandez2021} analysed the profitability of individual residential PV self-consumption in Spain, for the particular case of installations without storage. The conclusions reinforced the idea that grid-connected PV self-consumption is cost-effective in Spain under the new legal framework. It was also pointed out that collective self-consumption is one of the main pathways of improving economic profitability.

At the international level, several works reported the economic feasibility of self-consumption in the residential sector in a context of electricity market without subsidies \cite{Cucchiella2016,Lang2016,Bertsch2017,Camilo2017,Sagani2017}. The positive effect in self-consumption when including different strategies for electric vehicle charging was analysed in \cite{Kaschub2016}.
\cite{Schopfer2018} analysed how heterogeneity in real-world electricity load profiles affects the optimal system configuration and the cost-effectiveness of self-consumption systems in Zurich (Switzerland). Based on real energy consumption profiles, the authors found that system cost-effectiveness varied considerably between households, even for households with comparable total annual demand. While self-consumption without subsidies was profitable for 40\% of the households, batteries were not cost-effective in more than 99\% of the cases, unless storage prices dropped to a range of $250-500$ \euro/kWh. 
\cite{Espinoza2019} considered three real installations in Peru to analyse two different financial frameworks: one based on a lease contract and other considering a residential owner. While cost-competitiveness varied between the installations, the need for reducing the cost of capital was identified as a clear means to increase economic feasibility.

Based on the reviewed works, a number of issues and limitations when assessing the performance of self-consumption in households were identified. The following list includes a brief description of them, so that it can be useful for future researchers. A comment on how each of them is addressed in this paper is also included.

\begin{itemize}
    \item PV costs have experienced a sharp decrease in a relatively short period of time \cite{Creutzig2017,Haegel2019}. Since technology cost is a main driver of self-consumption cost-effectiveness, updated figures and sensitivity analyses are required. Table \ref{tab_PV_costs} gathers PV unitary costs assumed in the reviewed literature, broken down according to components where available.\footnote{Usually, an overall PV cost is reported (including inverter, structure, labour costs). Details provided on disaggregated costs are included in the last column of the table.} In this work, updated costs for small PV installations in Spain are employed (see details in Section \ref{sec_data_tool}), significantly lower than those referred in the table.

    \begin{table}[ht!]
        \centering
        \footnotesize
        \begin{tabular}{ccl}
        \hline
        Reference                   & PV unitary cost [\euro/kW]  & Comment \\
        \hline
        \cite{LopezProl2020}	     & $1,690$  & Residential \\
	\cite{LopezProl2017}        & $2,070$            &   Residential               \\ 
	\cite{Escobar2020}	     & $1,350$		    &  - \\	
        \cite{Cucchiella2016}       & $1,800 - 2,000$     &   Facility size 3-20 kW \\
        \cite{Lang2016}             & $2,090 - 3,330^{(a)}$  & Facility size $<$ 10 kW \\
        \cite{Bertsch2017}          & $1,130$   & $+1,330$ \euro \,  for installation     \\
        \cite{Camilo2017}           & $1,500 - 1,660$     &   Facility size 0.5-4 kW    \\
        \cite{Sagani2017}           & $1,700 - 1,900$     &   Facility size 2-10 kW   \\
        \cite{Kaschub2016}          & $1,600$    &  Inverter included (200-300 \euro/kW)  \\
        \cite{Schopfer2018}         & $2,000$              & -                          \\
        \cite{Christoforidis2016}   & $1,200 - 2,000$    & -          \\
        \cite{Sarasa-Maestro2016}   & $1,130$ &  PV:  550 \euro/kW, structure: 340 \euro/kW, inverter: 240 \euro/kW\\
         \hline
        \end{tabular} \\
        \footnotesize{$^{(a)}$ Conversion factor (2016): 0.9508 USD/\euro.}
        \caption{Unitary PV cost for self-consumption facilities considered in different works.}
        \label{tab_PV_costs}
    \end{table}

    \item Cost-effectiveness of self-consumption depends to a large extent on the optimal sizing of the installation \cite{Christoforidis2016}. \textit{Ad hoc} decisions on the design parameters may lead to suboptimal results and underestimations of the self-consumption potential. This could be the case in some works that assumed predefined parameters, such as PV capacities \cite{RoldanFernandez2021,Cucchiella2016,Lang2016,Camilo2017,Sagani2017} or the percentage of PV shared in consumption \cite{LopezProl2020, LopezProl2017,Cucchiella2016}.
    While it is true that PV capacities obtained from a techno-economic optimisation cannot be exactly matched by combining real-world available PV panels, in our opinion this is a practical issue to be addressed in a real project. At the research level, the use of predefined arbitrary PV capacities should be avoided. In this analysis, the sizing of the installation is an output of a techno-economical optimisation.  

    \item Some of the reviewed analyses consider real-world self-consumption facilities \cite{Mir-Artigues2018,Gimeno2018,Talavera2014,Espinoza2019}.
    Conclusions obtained in these cases should be taken with care, as the considered case studies may not be representative of the majority of potential self-consumers.
    In this work, average household buildings are obtained for every region in Spain. Each average building is characterised by the average number of households, floors, annual electricity consumption and rooftop surface (see \ref{subsec_databases}).
    
    \item Another issue in self-consumption analysis is the availability of data, specially load profiles and electricity price profiles. Load profiles are key in determining the overlapping between solar generation and electricity consumption, and the exchange dynamics with the grid. 
    According to the review included in \cite{Schopfer2018}, in most cases a single or a few number of measured household demand profiles are considered.
    Concerning electricity prices, a constant value related to average market price is often assumed for every hour of the year \cite{LopezProl2020, Mir-Artigues2018,Lang2016,Bertsch2017}. 
    This could be realistic in some places where retail prices are constant or vary according to a reduced number of periods. 
    However, this situation is likely to change completely in the following years. Generally accepted guidelines for electricity markets in a context of high penetration of renewables agree on translating price signals to consumers in order to promote demand-side management. In Spain it is possible for retail consumers to pay the electricity according to the hourly electricity market. In this work, hourly electricity prices and the best estimate of the load profile for households with hourly resolution for a whole year, which is published by the Spanish Transmission System Operator (TSO), are employed (see \ref{subsec_databases} for details). 
\end{itemize}

\section{Overview of current regulatory framework in Spain} \label{sec_legal_framework}

Up to April 2019, the deployment of self-consumption installations was seriously hindered in Spain. The initial and prolonged lack of legislation led out to a regulatory scheme that discouraged self-consumption by establishing very restrictive and economically detrimental conditions for such installations, contrary to what happened in several other countries \cite{IEA_review}. Finally, the recent enactment of Royal Decree 244/2019 (RD 244 from hereon) \cite{RD244} has established a regulatory framework that enables the deployment of domestic rooftop PV installations in the country.  The new legislation eliminates the sadly famous ``Sun tax'', \textit{i.e}., the tax imposed on the PV generation that is instantaneously consumed locally. Moreover, under RD 244, energy exported to the grid is economically rewarded and self-consumption installations shared by several consumers are allowed. \\

The current regulation identifies two kinds of installations, those direct selling electricity to the wholesale market and those with a simplified net billing system. Installations whose capacity is higher than 100 kW belong to the first category, and their electricity production must be sold to the wholesale market. This paper focuses on installations subject to a simplified net billing system, including or not compensation for the exported electricity.  They comprise household installations with a few kilowatts capacity, as well as larger systems supplying electricity to services or industrial buildings.\\

On the one hand, the owners of a self-consumption installation pay the electricity imported from the grid, at times when PV generation is lower than demand, at the usual price. In Spain, consumers can select either to sign a contract with a private electricity retailer or select one of the designated companies with a government-fixed price (PVPC, Voluntary Price for Small Consumers). On the other hand, the electricity exported to the grid, at times when PV generation exceeds local demand, is rewarded at a price that depends on the wholesale market price, as described in Section \ref{sec_remuneration_proposal}. Every month, the retailer charges the consumer the resulting net amount, together with capacity-dependent costs and taxes. Net balance cannot be negative. If the remuneration for the electricity exported to the grid is greater than the cost of the imported electricity, balance is zero and, in practice, the consumer is giving away the excess of generation at no cost. \\
 
Additionally, RD 244 allows nearby consumers to share one single installation. Shared installations must fulfil any of the following conditions: consumers shall be connected to the same low-voltage grid (downstream the MV/LV transformer), the distance between consumers' properties shall be lower than 500 m, or their registry numbers must share the initial 14 digits. Consumers within a shared installation must select fixed-in-time sharing coefficients\footnote{The Spanish legislator has announced that this will be modified soon to allow time-dependent sharing coefficients. This could be beneficial for shared installations among consumers with very different consumption patterns such as small businesses and households.}, which determine how the PV generation is distributed among them. As a final remark, the current legislation establishes that, for any kind of self-consumption installation, the user and the owner of the facility might be a different natural or legal person.

\section{Data and methodology} \label{sec_data_tool}

The analysis presented in this work combines different data from a number of databases. These data are inputs for the Distributed Energy Resources Customer Adoption Model (DER-CAM), a computational tool for techno-economical optimisation. DER-CAM determines the parameters of the optimal self-consumption installation under certain constraints, and generates the main performance indices. 

\subsection{The optimisation tool: DER-CAM}

The Distributed Energy Resources Customer Adoption Model, developed by Lawrence Berkeley National Laboratory (LBNL), is one of the world’s most popular software tools for economic planning of Distributed Energy Resources (DERs) in behind-the-meter and microgrid environments \cite{Stadler2014,Armendariz2017}.\footnote{The original version of DER-CAM can be accessed at no cost in: \url{https://building-microgrid.lbl.gov/projects/der-cam}}

Considering the specific conditions of a building (location, weather information, electric load, etc.) DER-CAM is able to calculate the optimal portfolio of DER investments (photovoltaic, diesel generators, storage technologies, etc.) that minimises the overall energy costs for the consumer. In this economic optimisation, DER-CAM takes into account techno-economic data of distributed generation technologies (including capital costs, operation, and maintenance costs; electric efficiency; maximum operating hours among others) as well as electricity tariffs and fuel costs. Due to the difference of lifetimes across DER technologies, 
the equivalent annual cost (\textit{EAC}), including annualised investment costs and operation costs, is the objective function that is minimised.
Batteries are operated in a daily cycle, by imposing that the state of charge at the end of the day is the same as in the beginning of the day.  
The optimisation problem is formulated assuming perfect foresight for the input data during the following day, which is a reasonable hypothesis, provided that electricity prices are known the day before, and solar radiation can be reasonable predicted.

For the purpose of the analysis of this paper, a DER-CAM version focusing on PV and storage technologies, and capturing the impact of battery ageing, was used \cite{Cardoso2018}. 
The investments in DER capacity (i.e. PV and storage) are disaggregated from the investments in power electronic components, such the battery controller and the inverter. Additionally, specific behind the meter constraints modelling the current regulatory framework imposed by RD 244 in Spain were introduced (in particular, a constraint limiting the amount of exported electricity to the grid that is economically compensated, see Section \ref{sec_remuneration_proposal}).

\subsection{Databases} \label{subsec_databases}

\begin{figure}[ht!]
    \centering
    \includegraphics[width=15cm]{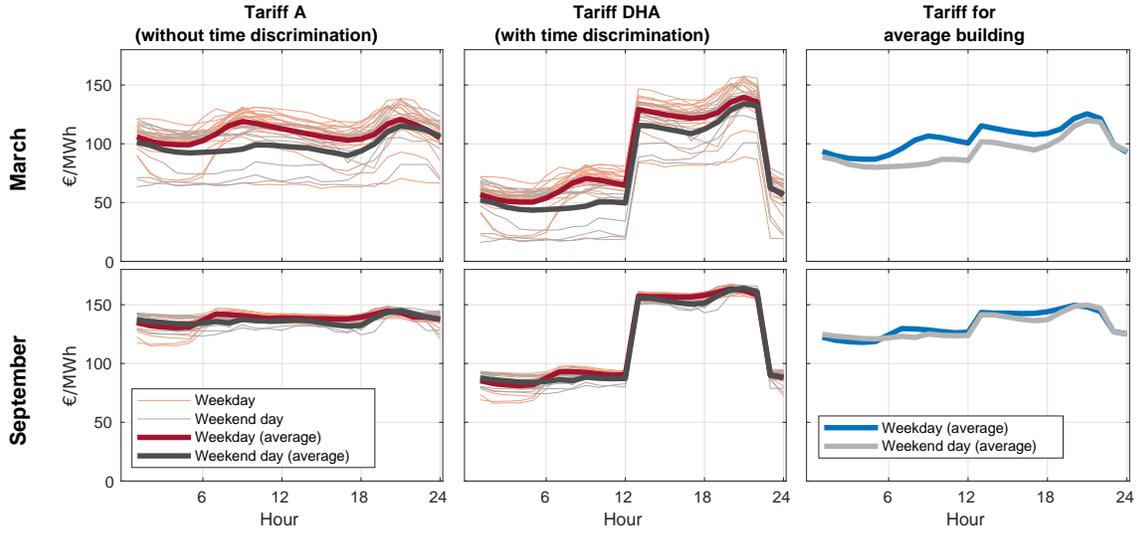}
    \caption{Electricity price given by PVPC tariff, considered for March and September, for options A (without time discrimination), DHA (with time discrimination) and for the average building. The latter is estimated based on the proportion of consumers in low voltage with discrimination options A, DHA and DHS, detailed in the text.}
    \label{fig_tariffs} 
\end{figure}

\begin{figure}[ht!]
    \centering
    \includegraphics[width=15cm]{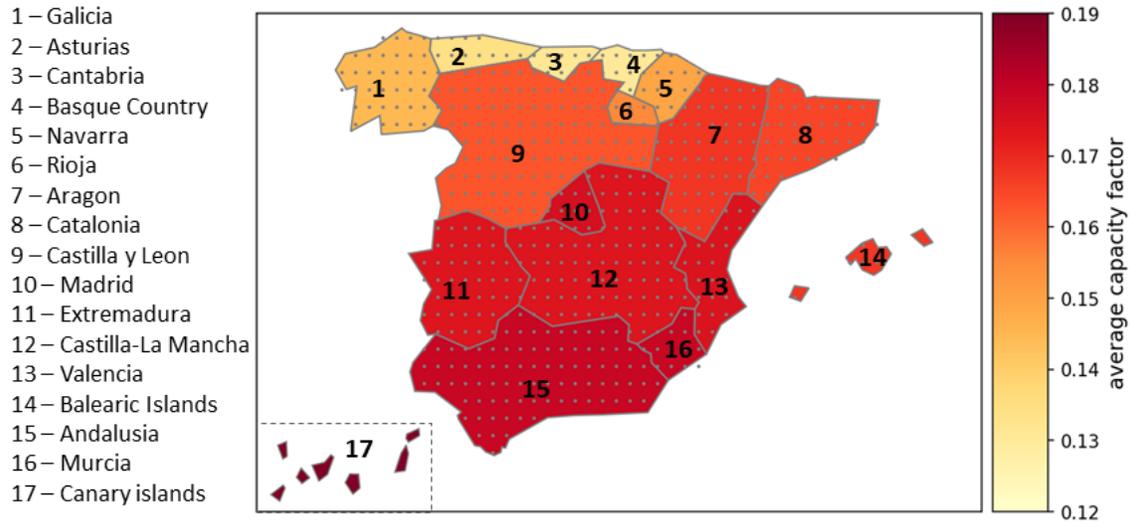}
    \caption{Capacity factors representative for every region in Spain. Average values over the 39 years of available data are shown. Grey dots indicate the spatial resolution of the reanalysis database from which solar radiation is retrieved.}
    \label{fig_02_solar_resource} 
\end{figure}

The different databases employed in the analysis represent the most up-to-date relevant data available at the moment for research purposes. These are:

\begin{itemize}
    \item Population and housing census (2011) \cite{INE2011}. This census is conducted in Spain every ten years by the Spanish Statistical Office. 
    Two relevant parameters have been derived from this database: average building floor area, $S$, and average number of households per building, $H$, at the regional level, as well as the national average, Table \ref{tab_data}. 
    \item Data from electricity consumers published by the Spanish independent regulator organism (CNMC) \cite{CNMC} and the Spanish institute for energy efficiency (IDAE) \cite{IDAE_Consumos}. These data have been employed to estimate:
        \begin{itemize}
            \item The average annual electricity consumption per household for every region in Spain, $C$, Table \ref{tab_data}.
            \item The share of consumers at a national level with the different time discrimination options: A (no time discrimination): $74.74\%$; DHA (two periods): $25.19\%$; and DHS (three periods): $0.07\%$.
        \end{itemize}
    \item Hourly profiles of final electricity consumption (2018) \cite{REE}. These profiles are published by Red El\'ectrica de Espa\~na (REE), the Spanish Transmission System Operator (TSO). They represent the best estimation on how the electricity consumption is distributed over the different hours of the year for consumers in low voltage, for the different time discrimination options. In this work, profiles for consumers with a contracted power below 15 kW were 
    employed to generate the daily load profile of the average building for week and weekend days, for every  month, and for every region in Spain. 
    \item Hourly electricity prices for consumers (Voluntary Price for Small Consumers, PVPC) for 2018, published by REE \cite{REE_ESIOS}. Under the PVPC scheme, the hourly electricity market price is directly translated to the consumer. Figure \ref{fig_tariffs} shows the PVPC for two months and for consumers without discrimination (A), two-period discrimination (DHA), and the averaged profile for the energy community (the average building). The PVPC includes the energy component of the access tariff, as described later in Figure 7. 

    \item Solar radiation time series have been produced using the Climate Forecast System Reanalysis (CFSR) database \cite{CFSR}, as described in \cite{Victoria_2019}. CFSR database includes hourly resolution and $40 \times 40$ km$^2$ spatial resolution. For every region in Spain, grid cells within the region are averaged to obtain representative time series. Figure \ref{fig_02_solar_resource} shows the annual capacity factor for every region (averaged over the 39 years of available data).  Hourly time series are used to create a representative day for every month that is used as input for the optimisation. 
    
\end{itemize}

\begin{table}[ht!]
    \centering
    \footnotesize
    \begin{tabular}{l|c|c|c}
\hline    
\textbf{Region}      & $S$ (m$^2$) & $H$ & $C$ (kWh) \\
\hline
Galicia             &   ~98.8   &   2.1     &   4,208  \\
Asturias            &   193.6   &   3.3     &   5,637  \\
Cantabria           &   197.0   &   3.2     &   3,922  \\
Basque Country      &   162.6   &   6.8     &   5,080  \\
Navarra             &   128.0   &   2.7     &   5,332  \\
Rioja               &   151.0   &   3.2     &   3,279  \\
Aragon              &   132.6   &   2.7     &   4,551  \\
Catalonia           &   120.2   &   3.4     &   4,074  \\
Castilla y León     &   ~92.6   &   2.0     &   3,246  \\
Madrid              &   167.8   &   5.3     &   2,933  \\
Extremadura         &   128.1   &   1.6     &   2,938  \\
Castilla la Mancha  &   116.2   &   1.6     &   3,080  \\
Valencia            &   131.2   &   3.1	    &   2,756  \\
Balearic Islands    &   114.3   &   2.4     &   3,203  \\
Andalusia           &   102.8   &   2.2     &   3,030  \\
Murcia              &   153.7   &   2.1     &   3,425  \\
Canary Islands      &   119.5   &   2.4     &   2,889  \\
\hline
NATIONAL            &   112.9   &   2.7     &   3,487  \\    
\hline
    \end{tabular}
    \caption{Regional and national average building. $S$: Average building floor area; $H$: Average households per building; $C$: Average household annual electricity consumption.}
    \label{tab_data}
\end{table}

\subsection{Cost assumptions}

Tables \ref{table_ref_data} and \ref{table_battery_data} contain the values adopted for a number of techno-economic parameters. Installation cost for rooftop PV system is assumed at 1,080 \euro/kW. This is similar to 1,127 \euro/kW estimated in Vartainen \textit{et al.} \cite{Vartiainen2017}, 1,100 \euro/kW shown for Germany in  \cite{Fraunhofer}, and 1,070 \euro/kW estimated for 2020 in the Danish Energy Agency Database \cite{DEA}. Out of the total cost, 1/3 (360 \euro/kW) is assumed for the modules and 2/3 (720 \euro/kW) for the Balance of System. In turn, Balance of System comprises the cost of the inverter estimated at 360 \euro/kW and the rest of elements (structure, wiring, labour) which are also estimated at 360 \euro/kW. The cost split is based on current prices for PV rooftop installations in Spain and roughly in agreement with the evolution of rooftop systems in Germany \cite{Fraunhofer}. The capacity of the inverter is set so that the DC/AC capacity is equal to 1.2 as this is the common practice when designing household PV systems \cite{handbook}.

Several battery technologies have been considered in order to devise the most suitable trade-off between cost and ageing parameter in the optimisation problem.
The ageing parameter of the battery represents the maximum number of cycles divided by the battery lifetime. The adopted values are based on \cite{Cardoso2018} and  \cite{Peters2017}.
The battery technology costs were considered to be between 335 and 667 \euro/kWh, depending on the technology. These values are in line with the current decreasing trajectories of storage technology costs \cite{Schmidt2017}.

Conservative assumptions have been made regarding the PV lifetime of 20 years, as many authors consider 25 years \cite{LopezProl2020,Camilo2017,Sagani2017,Espinoza2019,Christoforidis2016,Sarasa-Maestro2016} or even 30 \cite{Mir-Artigues2018,RoldanFernandez2021}. The discount rate was set according to the national debt funding costs at the moment of the research, 
which was 1.85\% for 15 years loan and 2.7\% for 30 years.
In our opinion, the use of a discount rate based on the WACC of the power energy sector in developed
countries (around 7\%) is not recommended as it reflects mostly the capital cost and the risk of
large-scale investments, such as grid-scale power plants. The case of self-consumption
investments, in particular at the household level, do not follow the same rationale because
the size of the self-consumption investments cannot be compared with the average of the
sector. The main reason for that is that the main investment alternatives to PV self-consumption, from the
perspective of individual citizens, are not investments in other power energy infrastructure, but
instead standard financial products offered by commercial banks. Similar figures can be found in \cite{Escobar2020} (3.5\%) and \cite{LaMonaca2017} (0.55\%).
Sensitivity analyses on both PV lifetime and discount rate are performed in Section \ref{sec_sensitivity}.

\begin{table}[ht!]
    \centering
    \footnotesize
    \begin{tabular}{llc}
        \hline
        PV                  &   Panel                   &   360 \euro/kW        \\
                            &   Installation+labour costs           &   360 \euro/kW        \\
                            &   Lifetime                &   20 years            \\
                            &   Rooftop occupation factor  &   10 m$^2$/kW         \\
        \hline
        Inverter            &   Unitary cost            &   360 \euro/kW        \\
                            &   Efficiency              &   0.93                \\ 
                            &   Lifetime                &   10 years            \\
        \hline
        Battery             &   Unitary cost            &   See Table \ref{table_battery_data}\\
                            &   Ageing parameter        &   See Table \ref{table_battery_data}\\
                            &   Lifetime                &   8 years             \\
                            &   Minimum state of charge &   5\%                 \\
                            &   Max. Charging rate      &   0.5 hours$^{-1}$   \\
                            &   Max. Discharging rate   &   0.5 hours$^{-1}$   \\
                            &   Efficiency (charge)     &   0.95                \\
                            &   Efficiency (discharge)  &   0.95                \\
        \hline
        Controller          &   Fix cost                &   50 \euro            \\
                            &   Unitary cost            &   60 \euro/kW             \\
                            &   Efficiency              &   0.95                \\
        \hline
        Other               &   Discount rate           &  2\%                  \\
        \hline
    \end{tabular}
    \caption{Techno-economic input parameters for the reference case.}
    \label{table_ref_data}
\end{table}

\begin{table}[ht!]
    \centering
    \footnotesize
    \begin{tabular}{llc}
        \hline
Type of battery                         &   Unitary   &   Ageing       \\
                                        &    cost     &    parameter      \\
\hline
{Lithium iron phosphate}                &	408 \euro/kWh &	562    \\
Lithium titanate		                &   668	\euro/kWh &	844   \\
Lithium Manganese Oxide		            &   335 \euro/kWh &	131   \\
Lithium Nickel Cobalt Aluminium Oxide	&	384 \euro/kWh &  394   \\
Lithium Nickel Manganese Cobalt Oxide	&	392 \euro/kWh &  225      \\
        \hline
    \end{tabular}
    \caption{Different types of batteries considered.}
    \label{table_battery_data}
\end{table}

\subsection{Performance indicators}

\begin{figure}[ht!]
    \centering
    \includegraphics[width=8cm]{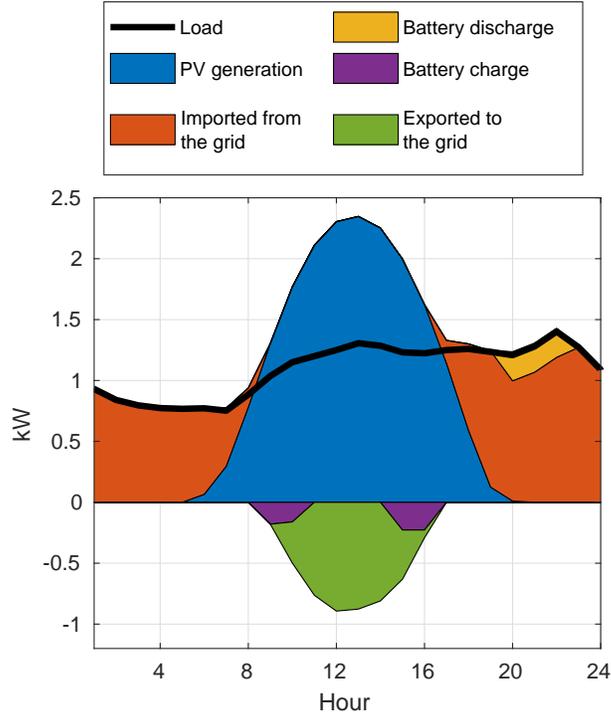}
    \caption{Example of one-day power flow computed with DER-CAM, for a self-consumption installation with storage and with remuneration for energy surplus.} \label{fig_powerflow_example}
\end{figure}

Results provided by DER-CAM include which elements are present in the installation (PV and/or storage), the optimal sizing that minimises \textit{EAC}, and the fraction of available rooftop occupied by the facility.
Time series with hourly power flow (\textit{i.e.} generated PV power, battery charge/discharge power, etc.) for week and week-end days for every month are also reported. Figure \ref{fig_powerflow_example} illustrates an example of the power flow for an installation including PV and batteries. The dispatch of storage output, storage input and PV sales is such that imported electricity occurs at minimum market price and exported electricity take place when remuneration is maximum (remuneration is related to market price, as described in Section \ref{sec_remuneration_proposal}; market price is known the day before).

The performance of the self-consumption installation is characterised by the following indices:

\begin{itemize}
    
    \item Annualised savings ratio (\textit{ASR}), defined as the percentage of reduction in \textit{EAC} compared to a case without self-consumption facility. Mathematically:   
    \begin{equation} \label{eq_ASR}
    \text{\textit{ASR}} = 100 \, \frac{\text{\textit{EAC}}_{noPV} - \text{\textit{EAC}}_{PV}}{\text{\textit{EAC}}_{noPV}} .
    \end{equation}
    It is worth noting that, for example, \textit{ASR}=5\% does not mean that consumers pay 5\% less in their electricity bill as a consequence of the initial investment required for the installation. It means that consumers pay 5\% less including both the electricity bills and the annualised initial investment. Thus, in practice, the percentage of reduction in the electricity bill would be higher.
        
    \item Self-sufficiency ratio (\textit{SSR}). 
    According to \cite{Luthander2015,Schopfer2018}, \textit{SSR} is defined as the percentage of annual load covered by PV and batteries, computed annually. 
    In Figure \ref{fig_powerflow_example}, \textit{SSR} is the PV generation below the load line plus the storage output divided by the total load. 
    100-\textit{SSR} represents the percentage of annual load covered with electricity imported from the grid.
    
    \item Self-consumption ratio (\textit{SCR}). 
    According to \cite{Luthander2015, Schopfer2018}, \textit{SCR} is defined as the share of generated solar energy that covers the load directly or via the battery. 
    In Figure \ref{fig_powerflow_example}, \textit{SCR} is the PV generation below the load line plus the storage output divided by the total PV generation. 
    100-\textit{SCR} represents the percentage of generated solar energy exported to the grid.

    \item Exported-imported ratio (\textit{EIR}), defined as the ratio between the energy exported to the grid and the energy imported from the grid.
    This ratio is important in order to check if an optimal installation is determined by the legal constraint regarding the amount of electricity that can be exported. Note that the ratio can be larger than 100\% (meaning that the amount of exported energy is larger than the imported energy), as the constraint refers the economic value of exported/imported energy (see Section \ref{sec_remuneration_proposal}).

\end{itemize}

For the cases not considering remuneration for the energy surplus, \textit{SCR} and \textit{EIR} will not be reported, as they take trivial values (100\% and 0\%, respectively) because there is no electricity injection into the grid.

\section{Regional results} \label{sec_regional_analysis}

The datasets and the computational tool described in Section \ref{sec_data_tool} have been employed to analyse the performance of the optimal self-consumption installation, including PV and storage technologies, for the average building in every region of Spain. The main results  are gathered in Table \ref{tab_regional} for installations without remuneration for the energy surplus and with remuneration, according to RD 244.

\begin{table}[ht!]
    \centering
    \footnotesize
    \begin{tabular}{lccccccccccr}
\hline 
       & \multicolumn{4}{c}{No remuneration} & & \multicolumn{6}{c}{With remuneration} \\
       \cline{2-5} \cline{7-12}
Region & PV/hh & Rooftop & \textit{ASR} & \textit{SSR} & & PV/hh & Rooftop & \textit{ASR} & \textit{SSR} & \textit{SCR} & \textit{EIR} \\ 
 & (kW) & (\%) & (\%) & (\%) & & (kW) & (\%) & (\%) & (\%) & (\%) & (\%) \\ 
\hline 
Galicia & 1.23 & 18.5 & 13.8 & 32.4 &  & 2.58 & 38.8 & 17.6 & 41.7 & 60.6 & 52.6 \\ 
Asturias & 1.69 & 19.9 & 12.2 & 31.3 &  & 2.88 & 33.9 & 14.9 & 39.3 & 72.3 & 27.4 \\ 
Cantabria & 1.17 & 13.2 & 11.3 & 30.4 &  & 1.97 & 22.1 & 13.8 & 38.5 & 73.6 & 24.8 \\ 
Basque Country & 1.53 & 44.9 & 11.2 & 30.5 &  & 2.52 & 73.8 & 13.5 & 38.3 & 75.0 & 22.8 \\ 
Navarra & 1.50 & 22.1 & 14.4 & 32.3 &  & 3.82 & 56.5 & 19.0 & 42.7 & 52.3 & 78.4 \\ 
Rioja & 0.94 & 14.0 & 15.4 & 33.6 &  & 2.63 & 39.4 & 21.1 & 43.6 & 46.7 & 103.7 \\ 
Aragon & 1.26 & 17.8 & 17.4 & 35.1 &  & 3.53 & 50.2 & 25.7 & 44.0 & 44.9 & 114.2 \\ 
Catalonia & 1.13 & 22.1 & 17.0 & 34.9 &  & 3.16 & 61.8 & 24.5 & 43.8 & 45.4 & 110.9 \\ 
C. Leon & 0.89 & 13.3 & 16.3 & 33.7 &  & 2.43 & 36.4 & 23.8 & 43.8 & 47.5 & 100.7 \\ 
Madrid & 0.78 & 17.2 & 18.0 & 35.0 &  & 2.16 & 47.3 & 27.5 & 44.1 & 45.7 & 110.9 \\ 
Extremadura & 0.81 & 06.9 & 18.3 & 35.8 &  & 2.24 & 19.1 & 27.6 & 44.5 & 44.8 & 116.9 \\ 
C. Mancha & 0.83 & 08.1 & 18.1 & 35.3 &  & 2.37 & 23.0 & 27.4 & 44.4 & 44.6 & 117.7 \\ 
Valencia & 0.75 & 12.4 & 18.3 & 35.8 &  & 2.10 & 34.5 & 27.2 & 44.4 & 44.9 & 116.0 \\ 
Balearic islands & 0.87 & 12.6 & 16.3 & 33.8 &  & 2.37 & 34.2 & 24.0 & 43.4 & 47.0 & 101.5 \\ 
Andalusia & 0.83 & 12.4 & 18.8 & 36.1 &  & 2.28 & 34.1 & 28.6 & 44.6 & 44.6 & 118.1 \\ 
Murcia & 0.92 & 08.6 & 18.8 & 36.1 &  & 2.60 & 24.3 & 28.4 & 44.5 & 44.3 & 119.7 \\ 
Canary islands & 0.76 & 10.9 & 19.8 & 36.6 &  & 2.25 & 32.2 & 31.9 & 45.1 & 42.1 & 135.5 \\ 
\hline 
NATIONAL & 1.00 & 16.4 & 16.2 & 34.5 &  & 2.73 & 45.0 & 22.8 & 44.0 & 46.7 & 105.4 \\ 
\hline 
\end{tabular}
    \caption{Performance of optimal self-consumption installations for regional and national average buildings. PV/hh stands for the PV capacity installed per household.}
    \label{tab_regional}
\end{table}

Results show that self-consumption installations in household buildings entail economic savings for consumers in all the regions, as compared to a situation without self-consumption. It is also noticed that, under the considered assumptions, storage is not part of the optimal configuration in any region, regardless the solar compensation policy.  

Considering the case without compensation for energy surplus, the installed PV capacity is, on average, 1 kW per household. 
The optimal installation covers a reasonably low percentage of the building rooftop in most cases (less than 22\% in all regions, except for Basque Country, where the percentage scales up to 45\%), meaning that the installation would be compatible with the presence of antennas or lifts, and shadowed areas would be relatively easy to avoid.
The self-sufficiency ratio (\textit{SSR}) ranges from 30\% to 36\%. 
Since there is no remuneration for energy surplus, the optimal installation is sized so that the  energy surplus is negligible.
Concerning the Annualised Savings Ratio (\textit{ASR}), it ranges from 11\% to 19\%, depending on the region. A significant difference is observed between the northern regions with a lower solar resource (Galicia, Basque Country, Asturias and Cantabria), for which the average \textit{ASR} is of around 12\%, notably lower than the average of the rest of the regions (17.5\%). However, the PV penetration obtained in northern regions is still relevant ($SSR$ above 30\% in all cases). This means that the environmental and social positive effects of self-consumption could also be obtained in northern regions, but the smaller cost-effectiveness could reduce the intensity of private investments, provided that self-consumption installations require that almost all the investment is taken in the first year.

Under the surplus compensation scheme, the installed PV capacity increases notably, reaching, on average, 2.7 kW of PV capacity per household. The PV penetration increases accordingly, with \textit{SSR} of around 44\%, and the annualised savings scale up to near 23\% in average. In average, more than half of the generated PV electricity is exported, as indicated by the national \textit{SCR} ratio of 46.7\%. In many regions (with special emphasis in southern regions), the exported energy is larger than the imported energy (\textit{EIR}$>100\%$), suggesting that the sizing of the optimal installation could be determined by the legal constraint concerning the amount of electricity that can be exported monthly with compensation, see description in Section \ref{sec_legal_framework}.
The fraction of rooftop covered by the installation also increases, with the national average reaching around 45\%. 

\section{Sensitivity analyses} \label{sec_sensitivity}

The reference case analysed in Section \ref{sec_regional_analysis} assumes a number of technical and economical hypotheses that impact the obtained results. This section deals with how the optimal configuration, sizing and performance of a self-consumption facility vary according to several parameters, such as the technology costs, the discount rate and the PV panel lifetime. 
The analyses are performed for the national average building, and considering the case without energy surplus compensation and with compensation according to RD 244.

\subsection{PV and battery costs} \label{subsec_PV_batt}

The technology costs (for PV panels and batteries) is of interest to analyse, as the current and future decreasing trends  \cite{Vartiainen2017} may redefine the whole scenario for countries to implement energy transitions. For the case of PV panels, a significant decrease of costs have been experienced over the last decade, for example making PV one of the most competitive energy sources worldwide, and paving the way for a massification of small-scale installations under appropriate conditions. Regarding storage technologies, while a certain cost reduction has already been observed, efforts are being placed in reaching further decreases, especially due to the critical role that batteries could play in decarbonising mobility \cite{Schmidt2017,Kittner2017}.

In this analysis ``PV costs'' include PV panel, inverter and installation costs. The reason for this is to facilitate comparison with other works, where an aggregated cost is usually reported. 
The considered range in this sensitivity analysis goes from 600 \euro/kW to 1,450 \euro/kW.
Concerning battery costs, the considered range includes cost reductions from -40\% to -75\% with respect to the reference cost of \euro/kWh. This range was found to be the appropriate one so that batteries are included in some of the optimal configurations. Since the optimal battery selected in all the cases is the Lithium-Nickel-Cobalt-Aluminium-Oxide battery, the battery costs vary from 230 \euro/kWh to 96 \euro/kWh. This is coherent with previous works, that concluded that actual battery costs are too high to be considered as a cost-effective option \cite{Camilo2017,Schopfer2018,Merei2016}. In particular, \cite{Schopfer2018} predicted storage prices in the range 250-500 \euro/kWh in Switzerland for storage to become cost-effective, and \cite{Merei2016} concluded that storage prices should be below 200 \euro/kWh in Germany.

\subsubsection{Case 1: No compensation for energy surplus}

\begin{figure}[ht!]
    \centering
    \includegraphics[width=15cm]{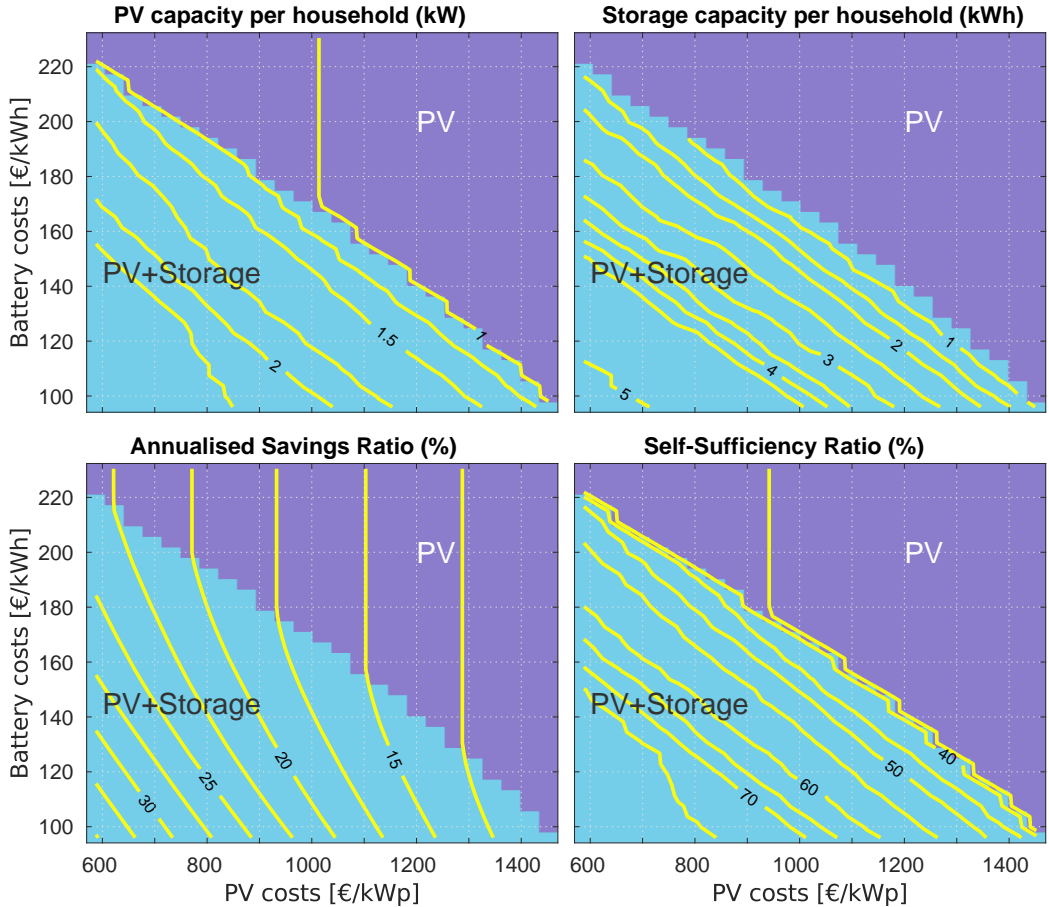}
    \caption{Sensitivity analysis of technology costs for case ``No compensation for energy surplus''. Purple and cyan areas indicate that the  optimal solutions comprise only PV panels, and PV panels plus batteries, respectively. Yellow iso-lines refer to the metric indicate in the title of the plot.} \label{fig_costs_case_1}
\end{figure}

Figure \ref{fig_costs_case_1} shows the different results obtained depending on the assumed PV and battery costs. 

Two areas can be observed in the plots presented in Figure \ref{fig_costs_case_1}, one corresponding to optimal installations including only PV panels (in purple) and other comprising PV panels and battery (in cyan). For the considered ranges of PV  and battery costs, self-consumption installations imply always annualised savings for consumers. Otherwise, a third region without PV panels and batteries would appear in the plots.

Interestingly, it is possible to observe that the installation of storage increases as the PV costs decrease. This is due to the fact that lower PV costs lead to high PV capacities and, therefore, more energy surplus. Since no remuneration for exported energy is considered in this case, the potential benefit of storage increases, making batteries more cost-effective. 

Focusing on the area of chart without storage, the optimal PV capacity is around 1 kW per household (as seen on the top-right panel), and it increases slightly when PV costs decrease. Consequently, the self-sufficient ratio slightly decreases (the installation
provides around 35\% of the annual electricity demand). Reductions in the PV panel costs affect mainly the annualised energy savings for consumers, which increase from 12\% to 22\% for the considered range of PV costs.

Regarding the area with storage, it is possible to observe a notably impact on the installation sizes caused by reductions in PV panels costs and/or reductions in battery costs. Indeed, an increase in both the optimal PV capacity and storage energy capacity can be observed, reaching more than 2.25 kW of PV per household, and a maximum of around 5 kWh of storage per household for very low technology costs. 
Self-sufficiency ratio increases accordingly, up to 80\%. In particular, results show that \textit{SSR} is increased by 1\% every -3 \euro/kWh in storage costs or every -18 \euro/kW in PV panels costs. 
It is also worth noting that rooftop occupation (not shown in the figure) also increases, reaching a maximum of 60\% for the considered cost ranges. Indeed, in an scenario of very-low technology costs, optimal self-consumption installations would require an extensive use of the available rooftop. 
Concerning economic performance, annualised savings also increase notably (above 30\%).

\subsubsection{Case 2: Compensation for energy surplus, according to RD 244}

\begin{figure}[ht!]
    \centering
    \includegraphics[width=15cm]{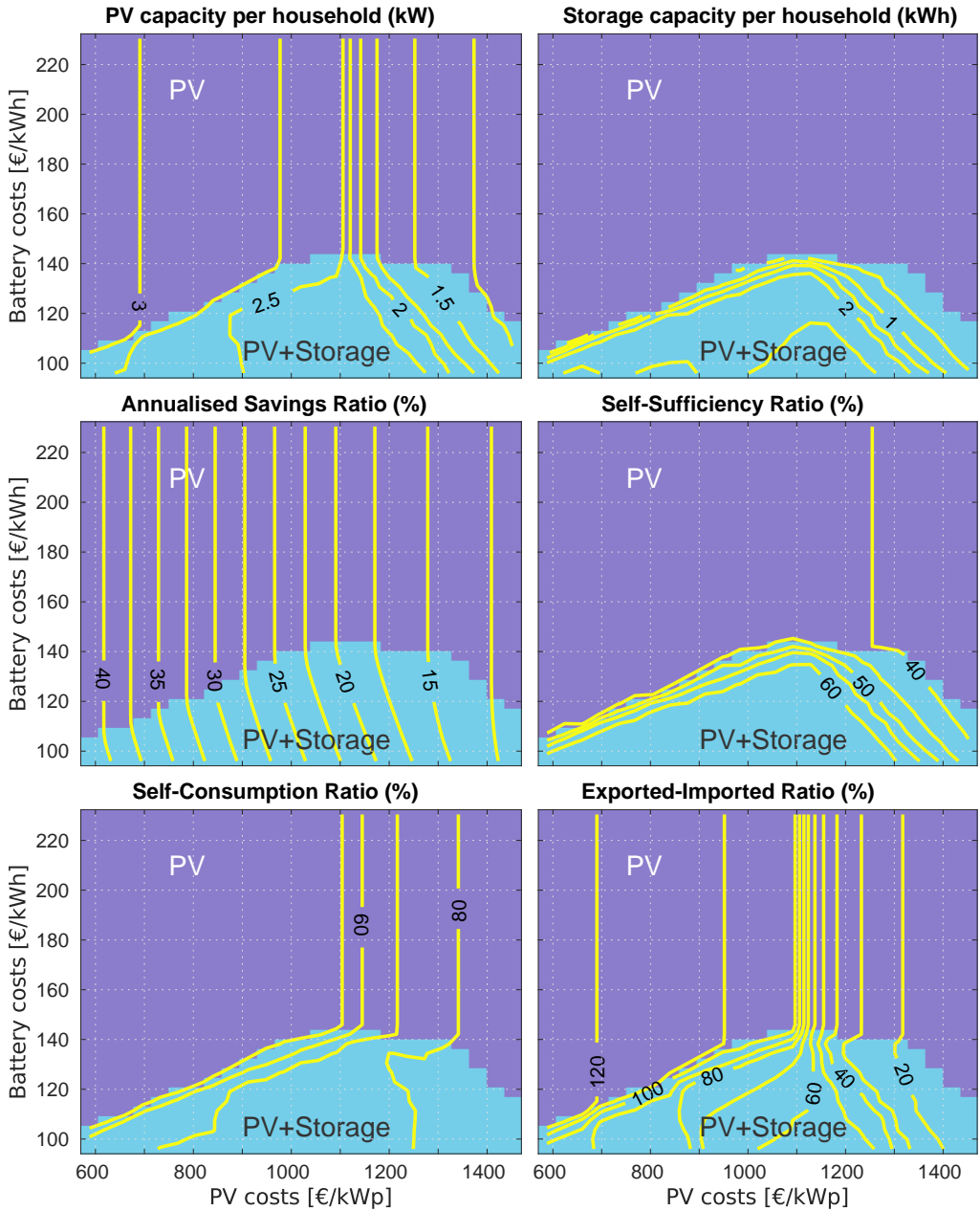}    
    \caption{Sensitivity analysis of technology costs for case ``With compensation for energy surplus''. Purple and cyan areas indicate that the  optimal solutions comprise only PV panels, and PV panels plus batteries, respectively. Yellow iso-lines refer to the metric indicate in the title of the plot.} \label{fig_costs_case_2}
\end{figure}

Figure \ref{fig_costs_case_2} shows the results obtained for different technology costs assuming a compensation for energy surplus.

As expected, for the same PV and battery costs, the annualised savings increase dramatically in comparison with the previous analysis, where no remuneration for energy exports was considered. The border between the two regions (with and without storage) provides interesting information regarding the impact of the remuneration policy on the configuration of the optimal installation, as discussed below.

Focusing on the area without storage, a decrease of the PV costs lead to a significant increase of the optimal PV capacity (from around 1 kW per household up to more than 3 kW per household). However, the obtained self-sufficiency ratios show that PV represents around 40\% of the electricity consumption, regardless the PV panel cost. 
This can be explained from the fact that the \textit{SSR}, in the absence of storage, is determined by the overlapping between PV generation and consumption. Therefore, after covering the demand, the additional PV capacity does not improve self-sufficiency, although it is still cost-effective due to the remuneration paid for the exports. 
As a result, lower PV costs decrease the self-consumption ratio (\textit{SCR}) and increase the exported-imported ratio (\textit{EIR}).
Interestingly, this effect is more pronounced for a range of PV costs from the maximum considered (1,450 \euro/kW) down to 1,100 \euro/kW, for which \textit{EIR} is below one. For PV panels cost lower than 1,100 \euro/kW, the increase of optimal PV capacity is less pronounced. This can be explained by the legislation constraints, which impose that, on a monthly basis, the economic value of the exported electricity cannot overcome that of the imported electricity. Since exported and imported electricity are valued with different prices (details provided in Section \ref{sec_remuneration_proposal}), it is possible to observe \textit{EIR} above 100\%. 

Considering the cases where storage becomes viable, two meaningful insights can be observed:
First, storage is only included in the optimal installation for battery costs below 140 \euro/kWh; in addition, this threshold varies with the assumed PV panel costs. 
Second, reductions in the PV panel costs favour the inclusion of batteries only in a PV cost range (above 1,100 \euro/kW). Further reductions in PV costs lead to lower storage adoption. 

This can be explained by looking at economics of the PV surplus, which can be either stored in a battery or fed into the grid. Storing in a battery implies consuming the PV surplus later at the value of the electricity rate, affected by the round-trip efficiency of the battery, \textit{i.e.} $0.95 \times 0.95 = 0.9$. In contrast, feeding PV into the grid implies selling the surplus immediately at a lower rate - as shown later in section \ref{sec_remuneration_proposal}, in 2018 this value was around 46\% of the energy rate (meaning that  for every kWh fed into the grid, the economic compensation allows for importing 0.46 kWh from the grid with no cost). 
This means that storage offers a higher value to the PV surplus than the grid (corresponding the difference between 0.9 and 0.46), but it implies an additional investment cost. 
Results show that, for low PV technology costs, these battery investments are not worthwhile and the PV surplus is valued at the grid export rate. 
Conversely, when PV technology costs increase, 
the most cost-effective configuration includes storage, as associated investments are compensated by the higher remuneration obtained for PV surplus.
Finally, as seen in the previous analysis, when PV technology costs are very high, the PV surplus is not cost-effective neither with storage nor with the grid rate, which means a decrease in both grid exports and storage investments.


\subsection{Sensitivity analysis of the discount rate} \label{subsec_interest_rate}

In this section, the sensitivity of the optimal installation with respect to the discount rate, $r$ is discussed. For renewable energy systems, the discount rate typically has a remarkable impact on the annualised energy cost (\textit{AEC}), as energy projects are characterised by high capital costs, $CC$, a long lifetime, $L$, and relatively low operational and maintenance costs, $OM$. These parameters are related according to the following expression:

\begin{equation} \label{eq_AEC}
AEC = CC \, \frac{r}{1-(1+r)^{-L}} + OM.
\end{equation}

Given this, considering for example $L=20$ years and $OM = 0$, the \textit{AEC} of a project corresponding to a scenario with $r=3\%$ is more than 20\% higher than the obtained with $r=1\%$. This gives an idea on the extent to which results can be affected by this parameter. \cite{Vartiainen2019} already highlighted the impact of discount rates for utility-scale PV installations.

\

For the case of a self-consumption installation, \textit{AEC} includes cash flows due to imports from/exports to the grid. For the case of an installation including PV and inverter but no  batteries, \textit{AEC} is given by:

\begin{equation}
    \text{\textit{AEC}} = C_{\text{\textit{imports}}} - R_{\text{\textit{exports}}} + CC_{\text{\textit{PV}}} \frac{r}{ 1-(1+r)^{-L_{\text{\textit{PV}}}}} + CC_{\text{\textit{Inv}}} \frac{r}{ 1-(1+r)^{-L_{\text{\textit{Inv}}}}},
\end{equation}

\noindent where $C_{\text{\textit{imports}}}$ is the annual cost associated with energy imports from the grid, and  $R_{\text{\textit{exports}}}$ is the remuneration due to exports to the grid. From this, it is clear that variations in the discount rate, $r$, are equivalent to changes in the capital costs of PV (panel+labour cost) and inverter, while $C_{\text{\textit{imports}}}$ and $R_{\text{\textit{exports}}}$ depend on the sizing of the installation (which in turn depends on PV and inverter capital costs) and the considered remuneration policy.

Thus, for a discount rate of $r = r_0 + \Delta r$, the equivalent capital costs with discount rate $r_0$ for PV and the inverter are given by:

\begin{equation}
    CC_{\text{\textit{PV,equiv}}} = CC_{\text{\textit{PV}}} \left(\frac{r}{r_0} \right) \frac{1-(1+r_0)^{-L_{\text{\textit{PV}}}}}{1-(1+r)^{-L_{\text{\textit{PV}}}}}, 
\end{equation}
 
\noindent and

\begin{equation}
    CC_{\text{\textit{Inv,equiv}}} = CC_{\text{\textit{Inv}}} \left( \frac{r}{r_0} \right) \frac{1-(1+r_0)^{-L_{\text{\textit{Inv}}}}}{1-(1+r)^{-L_{\text{\textit{Inv}}}}}, 
\end{equation}

\noindent respectively.

\

For the techno-economic parameters specified in Table \ref{table_ref_data} (in particular, $r_0 = 2\%$), different discount rates between 0\% and 5 \% are actually equivalent to keeping discount rate at 2\% while considering a total capital cost (PV+Inverter+Installation) between 860 \euro/kWp and 1,300 \euro/kWp. 
Thus, the same results obtained in Section \ref{subsec_PV_batt} for that cost range can also be attained by considering a discount rate between 0\% and 5\%. 

\

The high impact of the discount rate on the results indicates a possible strategy to incentive rooftop PV installations. If public institutions implement programs that include low-interest loans (even zero-interest loans), which will in practice reduce the discount rate, this would make rooftop PV systems more cost-effective and attractive for household consumers.

\subsection{Sensitivity analysis of the PV panel lifetime} \label{subsec_lifetime}

Following the reasoning detailed in Section \ref{subsec_interest_rate}, it is possible to relate variations in the PV panel lifetime ($L_{\text{\textit{PV}}} = L_{\text{\textit{PV,0}}} + \Delta L_{\text{\textit{PV}}}$) with an equivalent PV capital cost for the original lifetime ($L_{\text{\textit{PV,0}}}$). This equivalent cost is given by:

\begin{equation}
    CC_{\text{\textit{PV,equiv}}} = CC_{\text{\textit{PV}}}  \frac{1-(1+r)^{-L_{\text{\textit{PV,0}}}}}{1-(1+r)^{-L_{\text{\textit{PV}}}}}. 
\end{equation}

\

For the techno-economic parameters specified in Table \ref{table_ref_data} (in particular, $L_{\text{\textit{PV,0}}} = 20$ years), different PV panel lifetimes between 15 and 30 years are equivalent to keeping the life time at 20 years while considering a total capital cost (PV+Inverter+Installation) between 1,215 \euro/kWp and 825 \euro/kWp, respectively. The sizing and performance of the optimal installation for this range of total capital cost with a PV panel lifetime of 20 years was detailed in Section \ref{subsec_PV_batt} for the cases with and without remuneration for energy surplus.

\section{A proposal for upgrading the solar compensation mechanisms} \label{sec_remuneration_proposal}

In this section, the remuneration mechanism for exported energy described in the RD is analysed. As a consequence, a modification of this mechanism is proposed in order to reallocate the benefits due to avoided power losses.
It is important to stress that the object of this proposal is exclusively focused on the solar compensation mechanisms of the actual self-consumption legislation. 
Obviously, there might be other impacts (benefits and costs) related to the presence of the PV in the system and the solar compensation mechanisms are just a part of a broader discussion on rate design policy related to PV that involves: (i) assessing the technical impact of the PV in the system; (ii) quantify the added system costs or benefits that are related to the presence of  PV; (iii) allocate these costs or benefits to the tariffs, considering all type of consumers, and guarantee the cost neutrality of the rate design process. Such complex analysis, although important, is out of the scope of this paper.

As introduced in Section \ref{sec_legal_framework}, the new legal framework for self-consumption in Spain includes the possibility to remunerate the exported energy, according to a net billing mechanism. A number of different net billing schemes are currently implemented in different countries, following different criteria: Time-of-use tariffs (the remuneration tariffs are determined either based on historical data or in real time market prices), location-dependent tariffs (based on grid congestion at different nodes) and tariffs following avoided cost of electricity generation (avoided carbon emissions, avoided investments in transmission grid, avoided power losses, etc.) \cite{IRENA2019}.  
The new legal framework in Spain defines a net billing scheme based on market prices. The remuneration consists of a discount on the electricity bill, implemented in such a way that consumers pay for the difference between the \textit{economic value} of the energy consumed \textit{economic value} of the energy exported to the grid. This difference, on a monthly basis, cannot be negative.

In what follows, the case for self-consumers with the regulated tariff PVPC, as the case detailed in RD 244, is considered. Self-consumers in the free market have to establish private contracts with their retailers in order to define the remuneration mechanism. However, in our opinion, the mechanism described in RD 244 for the PVPC tariffs would set a reference frame for private retailers, who most likely would offer different versions of that mechanism, resulting in similar average remuneration prices. As a consequence, the mechanism for PVPC self-consumers would apply \textit{de facto} to all self-consumers.

The key point of the remuneration mechanism is that the referred \textit{economic value} is computed using differentiated prices for import and export energy, in which the latter is lower than the former. This is so because the definition of the remuneration price does not include some system cost components, for example, the energy losses due to electricity transmission and distribution. In order to understand the different prices employed to assess the economic value of imported and exported electricity, a brief description of the electricity price components in Spain is detailed below.

\begin{figure}[ht!]
    \centering
    \includegraphics[width=12cm]{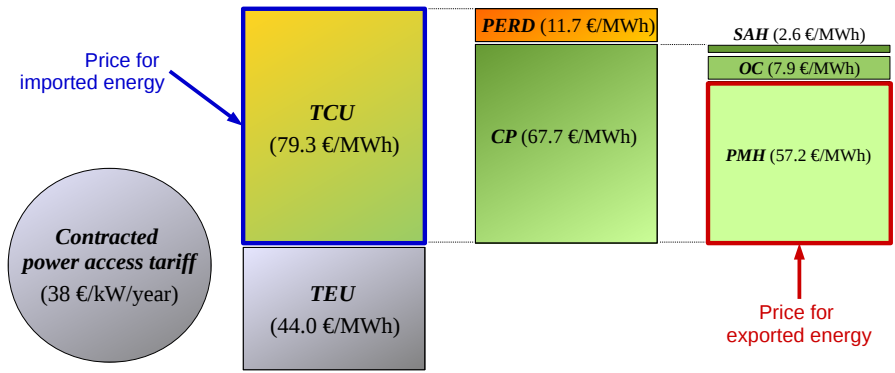}
    \caption{Components of the PVPC tariff depend on the contracted power (circle) and consumed energy (box). 
    \textit{TEU}: energy component of the access tariff.
    \textit{TCU}: energy production price. 
    \textit{CP}: production cost. 
    \textit{PERD}: power losses.
    \textit{PMH}: day and intra-day electricity market prices.
    \textit{SAH}: adjustment services cost.
    \textit{OC}: other costs.
    See text for additional details. In parentheses, average values for 2018 for consumers with no time discrimination tariff. Source: \cite{REE_ESIOS}.}
    \label{fig_04_PVPC} 
\end{figure}

Figure \ref{fig_04_PVPC} shows the PVPC tariff for consumers with no time discrimination, comprising a term that depends on the contracted power (in a circle) and other that depends on the consumed energy (in a box). The latter is broken down according to its different components. 
For those components whose value is time dependent, an average for 2018 is shown in parentheses. 
The access tariff (power and energy component) are shown in gray.

The energy term has two main components:\footnote{The acronyms are the official ones, and correspond to the acronyms in Spanish.} the energy component of the access tariff (\textit{TEU}), and the energy production price (\textit{TCU}). The \textit{TCU} component includes the production cost (\textit{CP}) and the cost associated to power losses (\textit{PERD}). The three components of \textit{CP} correspond to the day and intra-day electricity market prices, \textit{PMH}, the adjustment services costs, \textit{SAH}, and other costs ($OC$), including TSO and market operator remunerations and capacity payments, among others.
The component related to power losses, \textit{PERD} is computed as a percentage of \textit{CP}. This percentage used to be fixed to 14\% in the past \cite{SanchaGonzalo2014}, but it became variable every hour with the introduction of the PVPC tariff. Its average value in 2018 was above 17\%.

With the previous description in mind, the remuneration mechanism for the energy surplus described in RD 244 establishes that the economic value of the exported energy is computed using \textit{PMH}.
The average value of the \textit{PMH} in 2018 was $57.2$ \euro/MWh. The resulting economic value is subtracted from the economic value of the imported energy, for which \textit{TCU} is employed. The average value of \textit{TCU} in 2018 was $79.3$ \euro/MWh. In addition, the energy component of the access tariff, \textit{TEU}, is always added in the bill for the imported energy. The value of \textit{TEU} in 2018 was 44.0 \euro/MWh. Thus, this mechanism establishes the following two factors on the energy surplus remuneration (here quantified according to the average values of the electricity price components for 2018):
\begin{itemize}
    \item A constraint limiting the magnitude of the exported energy that is remunerated. For every kWh imported from the grid, the self-consumer is allowed to export up to $79.3/52.2 = 1.52$ kWh with remuneration (beyond that, exported energy is not remunerated).
    \item An assessment of the energy surplus remuneration. For every kWh injected into the grid, the economic compensation allows the self-consumer to take $57.2/(79.3+44.0) = 0.46$ kWh free from the grid in the same tariff period. 
\end{itemize}

The fact that \textit{SAH} and \textit{OC} components are not included in the remuneration price could be reasonable to some extent, as self-consumers benefit from the adjustment services incurred in the process of injecting energy and taking it back later (\textit{SAH}), and the other costs included in \textit{OC} are not avoided when injecting electricity surplus into the grid. 
Conversely, it is found that removing the \textit{PERD} component from the remuneration price is hardly justified. The energy surplus injected into the grid by a self-consumer, in a realistic scenario of distributed penetration of behind the meter PV, would be used by another consumer located nearby, avoiding conventional generation and, consequently, the associated losses due to energy transmission and distribution. However, the near consumer that takes this energy surplus from the grid does pay the \textit{PERD} component included in the tariff. This payment represents a benefit for the system, as it is not refunded to the self-consumer that injected the energy surplus because, according to RD 244, the \textit{PERD} component is excluded from the remuneration mechanism.  

An accurate energy surplus remuneration policy should properly reward investments that increase the efficiency of the system. In our opinion, this translates into including the value of the avoided power losses, given by the \textit{PERD} component, into the energy surplus remuneration. 
In order to assess the impact of this proposal, an analysis considering the following three remuneration policies for the energy surplus has been performed:

\begin{itemize}
    \item \textit{P1}: ``No remuneration''.
    \item \textit{P2}: ``Remuneration'', according to RD 244.
    \item \textit{P3}: ``Remuneration with losses'', proposed in this work, including the price component \textit{PERD} when computing the economic value of the energy surplus.
\end{itemize}

\begin{figure}[ht!]
    \centering
    \includegraphics[width=15cm]{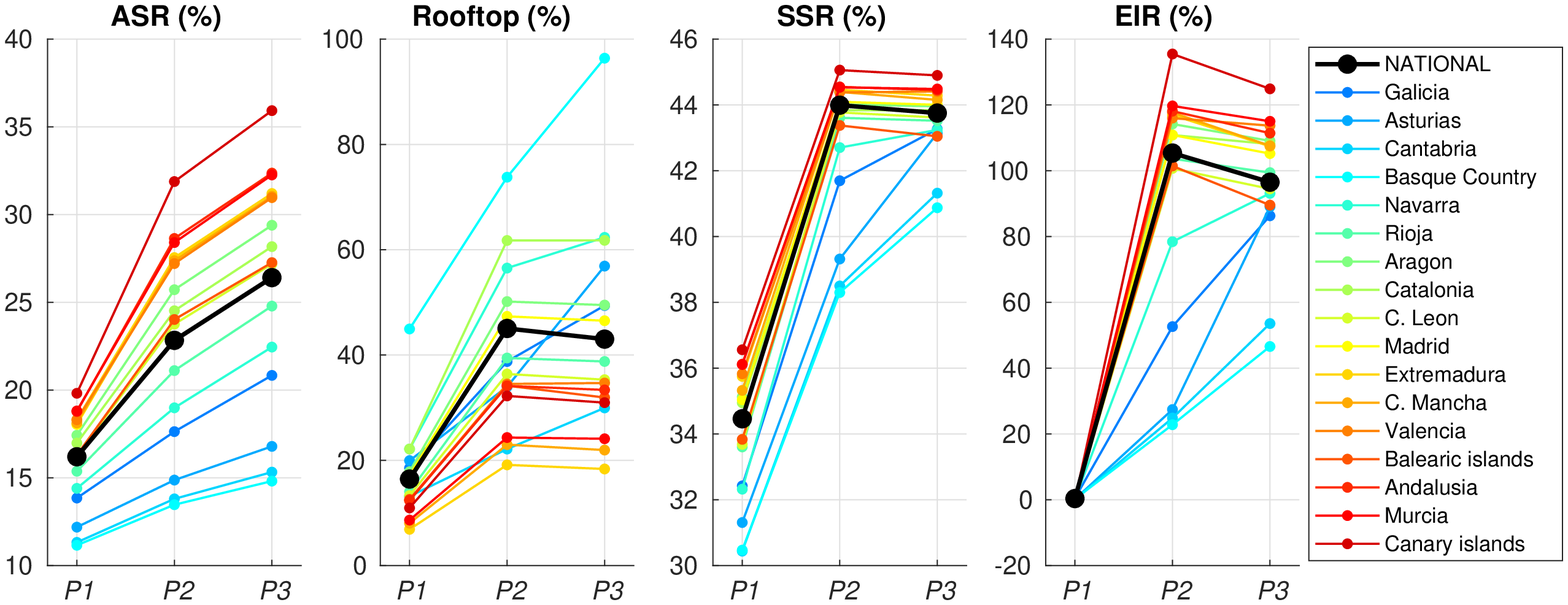}
    \caption{Regional performance of the optimal self-consumption installation according to different remuneration policies for energy surplus. \textit{P1}: No remuneration. \textit{P2}: Remuneration according to RD 244. \textit{P3}: Remuneration including the value of avoided power losses.}
    \label{fig_tarifas_PERD} 
\end{figure}

Figure \ref{fig_tarifas_PERD} shows the main performance parameters for the three considered remuneration policies, obtained for the optimal self-consumption installations in every region in Spain, and also for the average national building. 

From the economic point of view, as it could be expected, the proposed remuneration policy \textit{P3} would have a positive impact in all the territory. The Annualised Savings Ratio between policies \textit{P2} and \textit{P3} increases in around 3-4\% percentage points (except for the three northern regions, where the \textit{ASR} increases only around 1.5\%).

The other three plots in Figure \ref{fig_tarifas_PERD} reflect two sort of impacts of policy \textit{P3} in the different regions. For southern regions, the size of the optimal installation (thus, the occupied fraction of roof and the \textit{SSR}) remains constant (or decreases slightly). This is so because, for these regions, the size of the optimal installations under the remuneration policy \textit{P2} is determined by the legal constraint concerning the exported energy. Thus, even if the remuneration for energy excess is increased (policy \textit{P3}), the optimal PV capacity does not increase because additional energy surplus would not be remunerated. In fact, exports decrease a bit (see the plot on the right, \textit{EIR}) because the legal constraint makes reference to the economic value of exports, which is increased under policy \textit{P3}.
Consequently, for these regions, the proposed remuneration policy \textit{P3} has little impact for the optimal installation, but it has a positive impact from the economic point of view, as economic savings increases. It is worth noting that this increase in the annualised savings does not have extra cost to other consumers (self-consumer or not) or the administration. It comes as a mere consequence of a proper assessment of the benefits derived from reducing electricity losses.

Concerning northern regions, policy \textit{P3} leads to an increase in the optimal PV capacity.
Actually, both the \textit{SSR} and the \textit{EIR} increase notably. In a few cases, this increase in PV capacity leads to roof occupation above 50\% (with the particular case of Basque Country reaching 96\% of the average building roof surface). This means that, probably, in these regions, the optimal self-consumption facility could only be installed in a portion of buildings; this does not necessarily mean that self-consumption is not cost-effective, but that the installation performance would be below the optimal one as a consequence of limited rooftop availability.

\section{Conclusions and Policy Implications} \label{sec_conclusions}

In this work self-consumption in energy communities under the new legal framework in Spain has been analysed. A techno-economical optimisation performed with DER-CAM provided optimal installations for every region in Spain. A first contribution was the identification and the combination of relevant databases concerning load profile (week and weekend days), hourly electricity price (with and without time discrimination), regionally-resolved average building and solar resource. In addition, updated technology costs (household PV costs around 1,000 \euro/kW) and several battery technologies were considered.

Regional results showed that self-consumption is cost-effective in all the territory: savings in the annualised energy cost, as compared with a situation without self-consumption, ranged between 11-20\% for the case without remuneration for energy surplus, and 14-32\% with remuneration. The identified differences between northern and southern regions could be employed to define local policies in order to encompass self-consumption cost-effectiveness at national level.

Sensitivity analysis on technology costs revealed that batteries still require noticeably cost reductions to be included in the optimal self-consumption installation. In addition,
solar compensation mechanisms make batteries less attractive in a scenario of low PV costs, since feeding PV surplus into the grid, yet less efficient, becomes more cost-effective.

The high impact of the discount rate on the results suggested the need for specific policies (that could be implemented at national level or at the local level, taking into account the regional differences encountered in this work) oriented to help deploy self-consumption while ensuring fair conditions to small and new actors in the power system.

In our opinion, the reason why self-consumption promotion policies should be oriented to household consumers, with emphasis to those affected by energy poverty, is that regulatory frameworks must ensure fair conditions for small and new actors. Adequate financing schemes could be key to foster the needed emergence of distributed PV generation. 
Until now, power systems were based on large generation plants, whose required investments that could only be afforded by very large or state-owned companies. In Spain, as in many other countries, this led to the existence of very few large companies in the power sector. 
Distributed generation and self-consumption installations represent a paradigm shift, since they enable other stakeholders such as small businesses, cooperatives and citizens to play a role in the power system. The deployment of renewables in Germany \cite{yildiz_2014} and Denmark \cite{gorrono-albizu_2019} has resulted in a better distribution of generation property. Today, solar PV panels have reached such a low cost that distributed generation is competitive with other centralised technologies. However, as dominant actors in the power system try to retain their market power, changes in the current configuration of the system will not take place without public support.

An important aspect of the research was the analysis of the remuneration policy for energy surplus established in the new legal framework. A potential improvement was proposed, consisting in the inclusion of the value of the avoided power losses in the surplus remuneration. This upgrade of the legislation would allow for properly rewarding investments that increase the efficiency of the system.
However, this proposal needs to be framed in a broader discussion on optimal rate design, including an assessment of the technical impact of the PV in the system, a quantification of the added system costs or benefits that are related to the presence of PV, and the allocation of these costs or benefits to the tariffs, considering all type of consumers. 

The regional scale of the analysis allowed us to reveal that, in central and southern regions, the size of the optimal installation (for the case with remuneration) is determined by a legal constraint imposing limitations on the amount of exported energy that is remunerated. This is not the case for northern regions, for which the proposed upgrade of the remuneration policy would lead to higher PV optimal capacities and self-sufficiency rates.

Future lines of research include the impact of large-scale PV (and other RES) deployment on electricity market prices, allowing for a better performance assessment of self-consumption installations. In this line, future variations in consumption patterns should also be investigated and included in the analysis. Indeed, demand-side mechanisms are deemed to be key for the energy transition, and they may contribute to better integrate self-consumption facilities.

\section*{Acknowledgements}

This research was partially funded by the Lawrence Berkeley National Laboratory. The authors are thankful to this institution.

\bibliographystyle{unsrt}
\bibliography{bib_selfconsumption}   
\end{document}